\documentclass[%
reprint,
superscriptaddress,
 amsmath,amssymb,
prab,
]{revtex4-1}
\usepackage{siunitx}
\usepackage{subfigure}
\usepackage{graphicx}
\usepackage{dcolumn}
\usepackage{bm}
\usepackage[citecolor=none]{hyperref}
\usepackage{xcolor}

\begin{document}

\preprint{APS/123-QED}

\title{Demonstration of stable long-term operation of a kilohertz laser-plasma accelerator}

\author{L. Rovige}
\affiliation{ LOA, CNRS, \'Ecole Polytechnique, ENSTA Paris, Institut Polytechnique de Paris, Palaiseau, France}
\author{J. Huijts}%
 \affiliation{ LOA, CNRS, \'Ecole Polytechnique, ENSTA Paris, Institut Polytechnique de Paris, Palaiseau, France}
 \author{A. Vernier}
\affiliation{ LOA, CNRS, \'Ecole Polytechnique, ENSTA Paris, Institut Polytechnique de Paris, Palaiseau, France}
\author{V. Tomkus}
\affiliation{Center for Physical Sciences and Technology, Savanoriu Ave. 231, LT-02300, Vilnius, Lithuania}
\author{V. Girdauskas}
\affiliation{Center for Physical Sciences and Technology, Savanoriu Ave. 231, LT-02300, Vilnius, Lithuania}
\affiliation{Vytautas Magnus University, K.Donelaicio St. 58. LT-44248, Kaunas, Lithuania}
 \author{G. Raciukaitis}
 \affiliation{Center for Physical Sciences and Technology, Savanoriu Ave. 231, LT-02300, Vilnius, Lithuania}
 \author{J. Dudutis}
 \affiliation{Center for Physical Sciences and Technology, Savanoriu Ave. 231, LT-02300, Vilnius, Lithuania}
 \author{V. Stankevic}
 \affiliation{Center for Physical Sciences and Technology, Savanoriu Ave. 231, LT-02300, Vilnius, Lithuania}
 \author{P. Gecys}
 \affiliation{Center for Physical Sciences and Technology, Savanoriu Ave. 231, LT-02300, Vilnius, Lithuania}
 \author{M. Ouillé}
 \affiliation{ LOA, CNRS, \'Ecole Polytechnique, ENSTA Paris, Institut Polytechnique de Paris, Palaiseau, France}

\author{Z. Cheng}
\affiliation{ LOA, CNRS, \'Ecole Polytechnique, ENSTA Paris, Institut Polytechnique de Paris, Palaiseau, France}
\author{R. Lopez-Martens}
\affiliation{ LOA, CNRS, \'Ecole Polytechnique, ENSTA Paris, Institut Polytechnique de Paris, Palaiseau, France}
\author{J. Faure}
\affiliation{ LOA, CNRS, \'Ecole Polytechnique, ENSTA Paris, Institut Polytechnique de Paris, Palaiseau, France}

\date{\today}

\begin{abstract}
We report on the stable and continuous operation of a kilohertz laser-plasma accelerator. Electron bunches with 2.6\,pC charge and 2.5\,MeV peak energy were generated via injection and trapping in a downward plasma density ramp. This density transition was produced in a newly designed asymmetrically shocked gas nozzle. The reproducibility of the electron source was also assessed over a period of a week and found to be satisfactory with similar values of the beam charge and energy. These results show that the reproducibility and stability of the laser-plasma accelerator are greatly enhanced on the long-term scale when using a robust scheme for density gradient injection.
\end{abstract}

\maketitle


\section{\label{sec:intro}Introduction}

Laser-plasma wakefield acceleration \cite{tajima79} enables the generation and acceleration of electrons beams over very short distances due to their extreme longitudinal accelerating fields, orders of magnitude higher than in conventional accelerators. When driven by 100 TW to PW scale laser systems, Laser-Plasma Accelerators (LPA) can produce electron beams in the 100 MeV-GeV energy range and are being considered as drivers for femtosecond X-ray beams, either via betatron radiation \cite{Rousse2004} , Compton scattering \cite{TaPhuoc2012,Chen2013} , undulator radiation \cite{Fuchs2009} or free electron laser radiation. Such femtosecond X-ray beams could enable time-resolved (pump-probe) experiments based on e.g. X-ray diffraction or spectroscopy. A more recent line of research is currently focusing on the development of high-repetition rate (100 Hz-kHz) LPAs producing lower energy beams and requiring more modest laser parameters. TW-scale and kilohertz lasers with few-mJ pulse energy are capable of generating few MeV, pC range electrons beams \cite{Guenot2017,Gustas18,Salehi2017,faure2018} with femtosecond durations. Such beams could be used for low-energy applications such as ultrafast electron diffraction \cite{He2013,Faure2016} or irradiation of biological samples \cite{Rigaud2010,Lundh2012}.

In general, applications of LPAs require highly stable accelerator performances and the capacity to operate continuously and reliably over long periods of time, while providing consistent beam parameters from one day to the next. While numerous articles report on LPA stability on short time scales (as in a short series of shots over minutes) \cite{Faure2006,Osterhoff2009,Buck2013}, there is a lack of consistent data in the literature addressing the issue of long-term stability and reproducibility. This is presumably because most published studies have focused on proof-of-principle experiments and on studying the physics of various regimes of plasma acceleration. This is now changing with initiatives such as the LUX beamline at DESY, aiming at turning a LPA into an actual machine \cite{DELBOS2018}. In Ref. \cite{DELBOS2018}, day-long operation at a few Hz repetition rate is reported. However, to the best of our knowledge, no data has actually been published.\par
Concerning kHz laser-plasma accelerators, results obtained in the resonant condition for the bubble regime \cite{Guenot2017,Gustas18} yielded electron beams with pC charge in the MeV range. These experiments have reported on short term stability of the electron beam distribution but long-term stability has remained a challenge so far. This is because the kHz laser systems currently used to drive high-repetition LPAs have parameters that are at the limit of what is necessary to efficiently inject and accelerate electrons in the wakefield. In previous experiments, electron beam generation relied on ionization injection \cite{Pak2010,McGuffey2010,Guenot2017} and self-injection \cite{Gustas18}. Both mechanisms being rather sensitive to the laser intensity, this made electron injection in the wakefield difficult to control and stabilize, resulting in a high sensitivity to experimental parameters such as laser intensity, pulse duration and small laser misalignments. Therefore, long-term stability has been difficult to achieve in this regime.\par
On the other hand, controlled injection techniques are known to increase the reliability of electron injection and, consequently, beam stability \cite{Faure2006}. In particular, injection in a plasma density-transition \cite{Bulanov1998,Tomassini2003,Kim2004,Suk2004,Brantov2008} is relatively straightforward to implement experimentally. While the laser pulse goes through the density transition, the plasma wavelength increases so that the bubble is rapidly elongated, causing an effective slowing down of its back that facilitates the trapping of electrons. The concept of downward density ramp injection was validated in  proof-of-principle experiments \cite{Geddes2008,Faure2010}. Other experiments relying on shock-front injection have also been successful \cite{Schmid2010,Thaury2015,Swanson2017}. In these experiments, a sharp density transition is obtained by inserting a knife-edge into the supersonic flow of the gas jet. These results indicate that density transition injection consistently triggers  localized injection, and could provide the necessary stability for long-term and reproducible operation of a kilohertz laser-plasma accelerator. \par 
In this paper, we report on the continuous and autonomous operation over 5\,hours of a kilohertz LPA, generating MeV-pC range electron beams thanks to the use of a newly designed asymmetrically shocked supersonic gas jet allowing injection in a sharp downward density gradient. The paper is organized as follows: in section II, we first describe the experimental apparatus and in particular the gas jet design for generating a density transition. In part III we show the result of 5h operation and day-to-day reproducibility. Finally, we conclude in part III.

\section{\label{sec:results}Experimental methods}
\begin{figure}[h!]
\includegraphics[width=9cm]{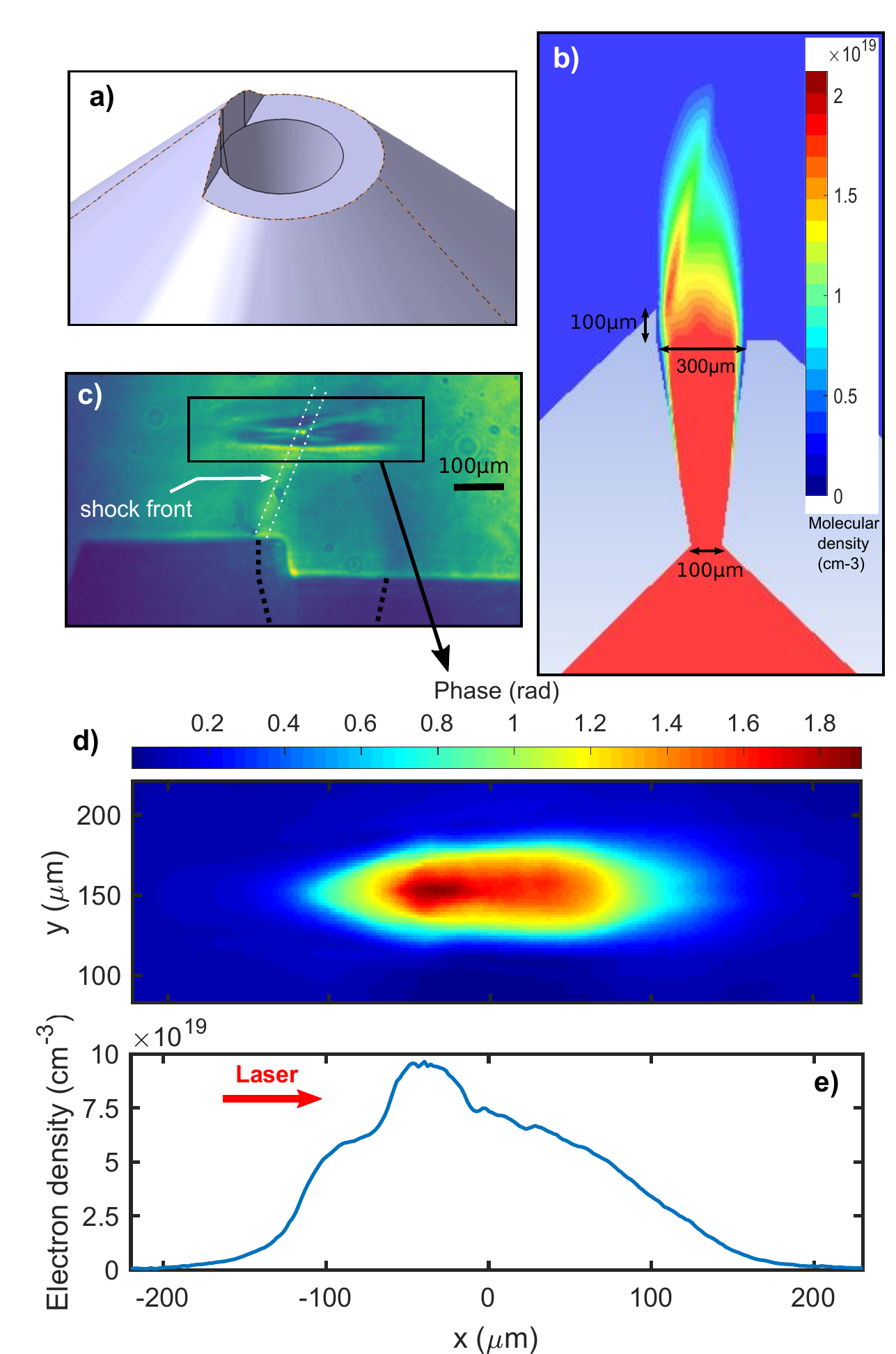}
\caption{\label{fig:jet}a) 3D model of the One-Sided Shock Nozzle. b) Molecular density map from a 3D CFD simulation of an OSS jet performed with Fluent. Gas is $\mathrm{N_2}$ and backing pressure P\,=\,20\,Bar. Colormap is capped at $\SI{2.1e19}{\per\square\centi\meter}$ for viewing purpose. c) Shadowgraphic side view image of the plasma. Black dotted line suggests the inner walls of the nozzle, white dotted lines follow the shock front. d) Experimental phase from the plasma column measured with the wave-front sensor. The $y$-axis represents the distance from the OSS nozzle's exit, the $x$-axis is the laser propagation axis. e)  Electronic density profile at $\SI{150}{\micro\meter}$ from the OSS nozzle's exit, retrieved by Abel inversion of the phase map. Laser propagation direction is from left to right (red arrow).}
\end{figure}

 \begin{figure*}[th!]
     \centering
         \includegraphics[width=16cm]{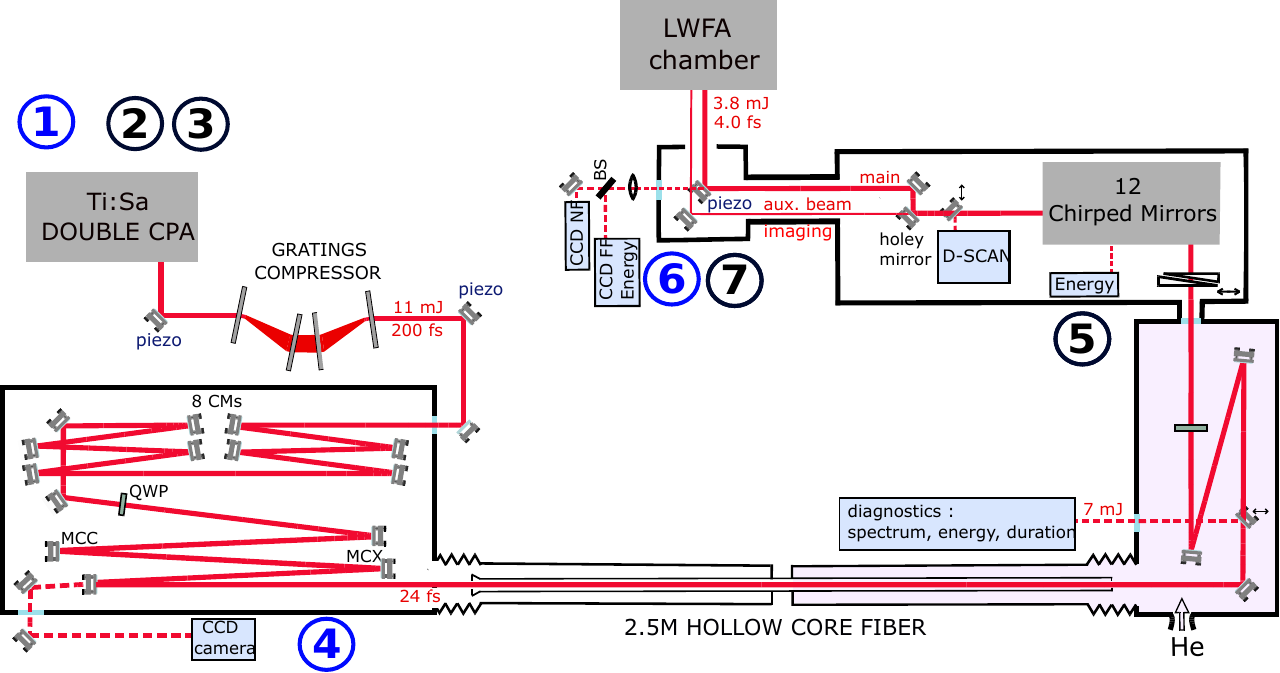}
        \caption{\label{fig:laser} Schematic of the laser system. Numbers \raisebox{.5pt}{\textcircled{\raisebox{-.9pt} {1}}},\raisebox{.5pt}{\textcircled{\raisebox{-.9pt} {4}}},\raisebox{.5pt}{\textcircled{\raisebox{-.9pt} {6}}} show the position of the beam pointing stabilization devices, numbers \raisebox{.5pt}{\textcircled{\raisebox{-.9pt} {2}}},\raisebox{.5pt}{\textcircled{\raisebox{-.9pt} {3}}},\raisebox{.5pt}{\textcircled{\raisebox{-.9pt} {5}}},\raisebox{.5pt}{\textcircled{\raisebox{-.9pt} {7}}} correspond to the position of energy diagnostics for the booster amplifier, the power amplifier, the Hollow Core Fiber (HCF) compressor and the turning box, respectively.}   
\end{figure*}

\subsection{One-sided shock nozzle}
Previous implementation of the density transition injection scheme relied on the lateral insertion of a knife-edge after the exit of the jet in order to induce a shock-front in the gas flow \cite{Schmid2010,Buck2013}. We propose a similar method, but the formation of the shock-front is directly incorporated in the design of the nozzle, which consists of a $\SI{100}{\micro\meter}$ throat and  $\SI{300}{\micro\meter}$ exit diameter ``De Laval" nozzle to which a $\SI{100}{\micro\meter}$ long flat section has been added at the end of one side of the nozzle (see Fig.~\ref{fig:jet}.a). These nozzles will thereafter be referred to as ``One-Sided Shock" (OSS) nozzles. The flat section induces an abrupt change of direction in the gas flow which, due to the flow being supersonic, translates into a shock-front formation and therefore leads to a sharp density gradient downstream \cite{Zucker2002}. Compared to inserting a knife edge into the jet, incorporating the shock-front generation directly in the nozzle design offers a solution that is more robust and easier to install, for such a small nozzle. However, machining such a complicated nozzle geometry, including micrometer scale features, is technologically demanding. In practice, the nozzle was made using nanosecond laser rear-side processing and the Femtosecond Laser-assisted Selective Etching technique (FLSE,\cite{Marcinkevicius2001,Tomkus18}) which consists of chemical etching of previously selectively laser-irradiated fused silica.  3D computational fluid dynamics (CFD) simulations were realized with the software Fluent to validate and optimize the design. Figure \ref{fig:jet}.b shows a  map of the molecular density obtained by simulating a nitrogen flow through the OSS nozzle, with a backing pressure of 20\,bar. The shock-front in gas density originating from the final straight section is clearly visible.\par
The plasma density profile was characterized experimentally by sending the laser pulse into a nitrogen gas jet produced from the OSS nozzle. The plasma column produced by the main beam was illuminated from the side by a probe beam, and imaged on a quadriwave lateral shearing interferometer (SID4-HR, Phasics, \cite{Primot1995,Primot2000}). The plasma density line outs can be derived from the phase maps (see Fig.~\ref{fig:jet}.d) via Abel inversion, assuming radial symmetry around the horizontal axis. Figure \ref{fig:jet}.e shows the measured electron density line out at $\SI{150}{\micro\meter}$ from the nozzle exit, with a backing pressure of 22\,bar. The peak density is $\SI{9.7e19}{\per\square\centi\meter}$ and the density after the shock is $\SI{7.3e19}{\per\square\centi\meter}$, corresponding to a 25\% density drop with a transition width of $\SI{15}{\micro\meter}$. Note that with nitrogen, a single molecule of $\mathrm{N_2}$ produces 10 electrons once ionized by the laser to $\mathrm{N^{5+}}$, which provides the high plasma density required to resonantly drive the plasma wakefield, while keeping the background pressure in the vacuum chamber to a reasonable level and therefore making continuous kilohertz operation possible. \par

\subsection{Experimental set-up}
The experiment was conducted using the Salle Noire laser system at LOA \cite{Bohle2014,Ouille2020}, which provides 10\,mJ, 25\,fs FWHM laser pulses at kilohertz repetition rate, with a central wavelength $\lambda_0=800\,\mathrm{nm}$. The pulse is then post-compressed in a helium-filled Hollow Core Fiber (HCF). Through a pair of motorized fused silica wedges we can control the amount of dispersion to fine-tune pulse compression or add chirp to the pulse, and measure the pulse temporal intensity profile using the d-scan technique \cite{Miranda2012}. The energy on target was 3.8\,mJ and the pulses are focused by a f/4 off-axis parabola, resulting in a $\SI{6.2}{\micro\meter}\times\SI{5.5}{\micro\meter}$ FWHM focal spot, which corresponds to a Rayleigh range of $z_R\sim \SI{100}{\micro\meter}$. These laser parameters yield a measured peak intensity in vacuum of $\mathrm{I}=\SI{2.0e18}{\watt\per\square\centi\meter}$ and a normalized vector potential  $\mathrm{a_0}\simeq1.0$.  The charge as well as the electron beam distribution were measured with a calibrated CsI(Tl) phosphor screen imaged onto a high dynamic range CCD camera. The laser beam was blocked by a thin aluminum foil in front of the phosphor screen, so as to block electrons with energies below 100\,keV. The energy of the electrons was measured with a retractable spectrometer made of a $\SI{500}{\micro\meter}$ pinhole and two permanent cylindrical magnets.\par
It is important to note that the laser-plasma accelerator truly runs at a repetition rate of 1 kHz: the gas jet flows continuously while the pumping system is able to maintain a vacuum of a few $10^{-3}$~mbar; the data was continuously collected during the experimental run, often by acquiring data accumulated over several shots. In order to perform a long run it was necessary to implement numerous diagnostics and feedback loops in the laser system to make sure that the laser parameters remained stable over a long time. Figure \ref{fig:laser} shows a schematic of the laser system with indicated positions where laser energy monitoring was performed. In order to maximize the stability of the energy, the spectrum, and the spatial mode output of the HCF, a fast beam pointing system is essential. In our case, it ensured an energy stability within one percent over the course of a full day. Considering the long optical path from the exit of the fiber to the experiment itself, compensation for long-term thermal drifts proved crucial to the long-term stability of the LPA. We thus set up a slow beam pointing device operating at $\mathrm{<1\,Hz}$ just before the experiment, using the leakage through of mirror in the turning box just before the final focusing parabola-- point {\textcircled{\raisebox{-.9pt} {7}}} in Fig.~\ref{fig:laser}. This ensured that the laser beam alignment on the gas jet stayed rigorously the same and that the focal spot quality was kept consistent throughout the long acquisition run.

\begin{figure*}[ht!]
     \centering
         \includegraphics[width=17cm]{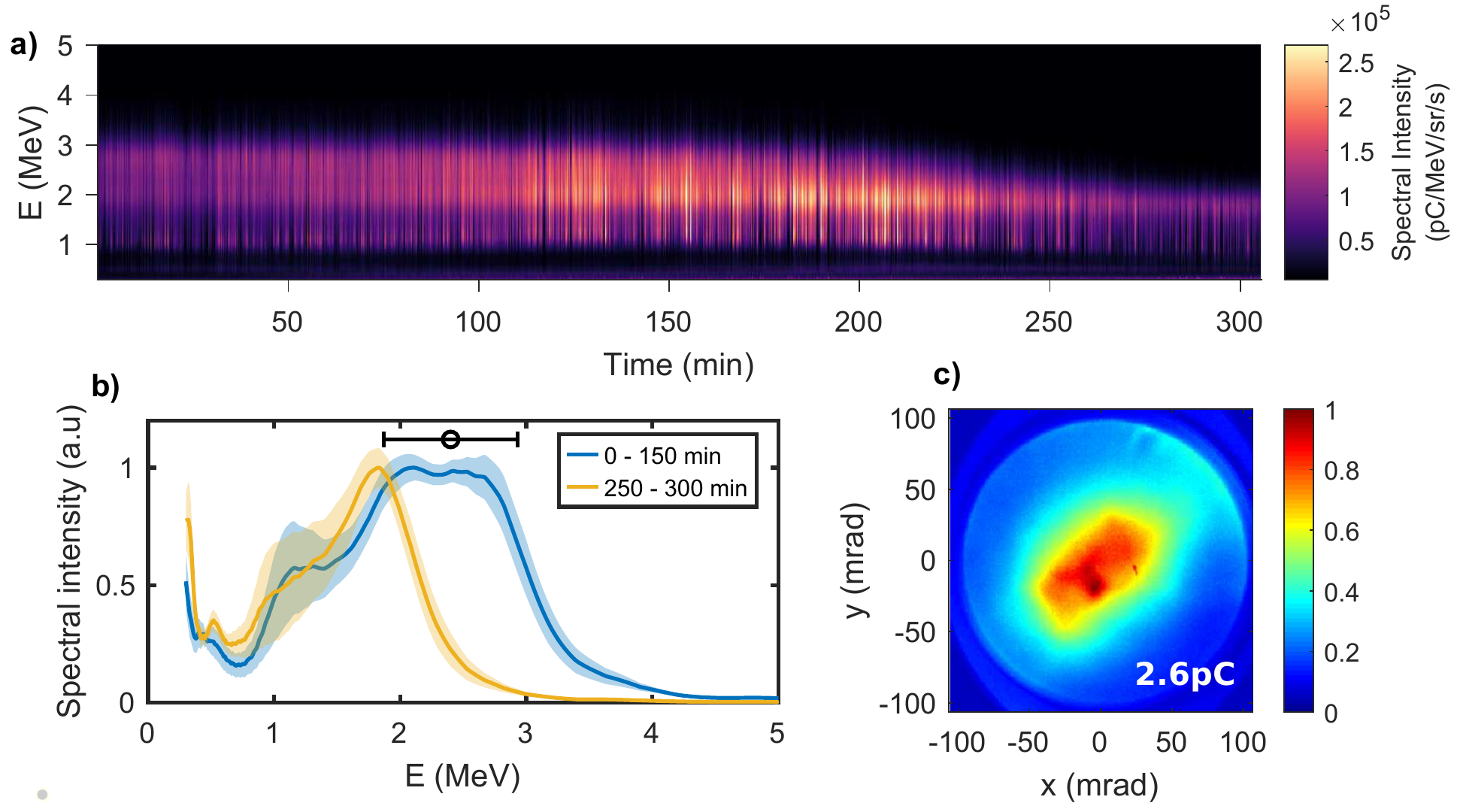}
        \caption{\label{fig:electrons} a) Electron spectra measured continuously for 306\,min. Each spectrum is averaged over 100\,shots. b) Blue line: mean spectrum over the 0-150\,min period. Yellow line: mean spectrum over the 250-300\,min period. Shaded areas show standard deviation. The black error bar indicates the spectrometer resolution at 2.4\,MeV. c) Electron beam measured just before the start of the 5h spectrum monitoring. The total charge per shot is $\SI{2.6}{\pico\coulomb} \pm \SI{0.6}{\pico\coulomb}$\,(std). The beam divergence is approximately\,$\SI{80}{\milli\radian}$ FWHM. }
\end{figure*}

\section{\label{sec:results}Experimental results}

\subsection{\label{sub:results}Stability over several hours of continuous operation }

The beam profile and charge were measured right before the start of the electron spectrum monitoring. Statistics were performed from 20 acquisitions, each consisting of an accumulation over 10 shots, thus accounting for 200\,shots in total. This initial measurement yielded a mean charge of $\SI{2.6}{\pico\coulomb}$ per shot with a 0.6\,pC standard deviation, and a beam divergence of 80$\times$75\,mrad FWHM (see Fig.~\ref{fig:electrons}.c). The electron spectrum was then monitored during 5 hours of complete hands-off operation of the kilohertz laser-plasma accelerator, i.e. with no other intervention than the beam pointing stabilization feedback loops at the three above-mentioned locations in the laser chain, see Fig.~\ref{fig:laser}. Results of this measurement are displayed in Fig.~\ref{fig:electrons}.a. Beams with peaked spectra and a large majority of electrons of energy greater than 1\,MeV were reliably produced throughout the whole 306\,min of monitoring.  Moreover, during the first 150\, min, the spectrum remained very stable, with a peak energy of 2.5\,MeV. After that, the high-energy part noticeably eroded with time, lowering the peak energy to 1.9\,MeV. Possible reasons for this energy loss will be addressed later in this article. A comparison of the spectra during the first 150\,min and the last 50\,min is represented in Fig.~\ref{fig:electrons}.b.\par

\begin{figure}[th!]
     \centering
         \includegraphics[width=8cm]{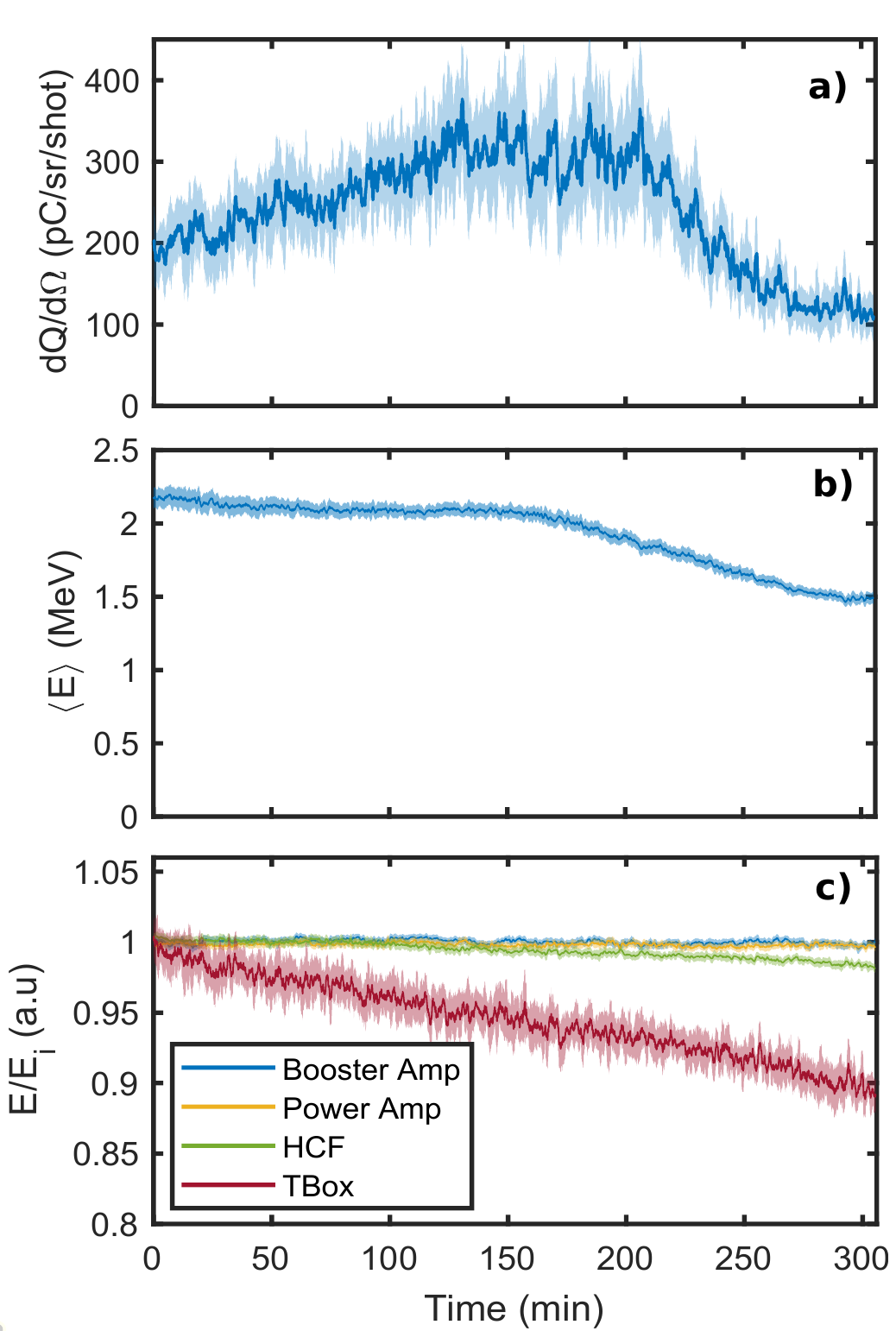}
        \caption{\label{fig:stab}a) Total charge per solid angle collected through the electron spectrometer pinhole. b) Mean electron beam energy versus time. c) Monitoring of the laser energy at four points in the laser chain. Each curve in this figure is averaged over a one minute moving window and shaded areas represent the corresponding standard deviation.}   
\end{figure}
\begin{figure}[h!]
     \centering
         \includegraphics[width=8cm]{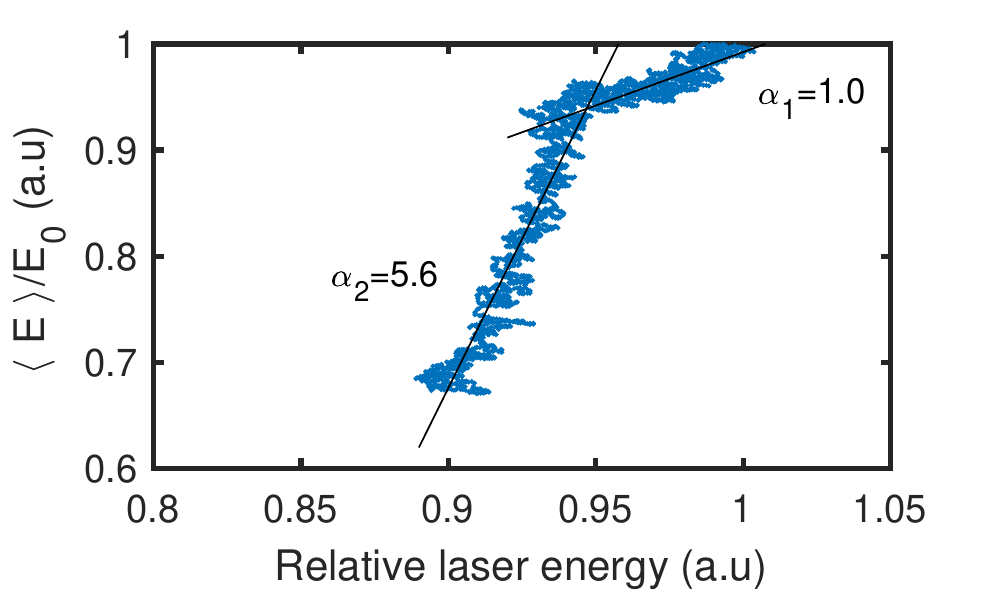}
        \caption{\label{fig:corr} Correlation between the electron mean energy and the relative laser energy at the turning box measurement point (right before the off-axis focusing parabola), and linear fits of the two different parts of the curve, with slopes $\mathrm{\alpha_1}$ and $\mathrm{\alpha_2}$.}   
\end{figure}

To complete these data and assess more thoroughly the question of stability, we plot the temporal evolution of the total charge per solid angle $dQ/d\Omega$ collected through the pinhole of the spectrometer  (Fig.~\ref{fig:stab}.a), the mean energy of the electrons $\langle E\rangle$ (Fig.~\ref{fig:stab}.b), as well as the relative laser energy at different points of the laser chain (Fig.~\ref{fig:stab}.c). All curves are averaged over a 1\,min moving window. 
The data show an increase in $dQ/d\Omega$ (Fig.~\ref{fig:stab}.a) during the first 130\,min. This is likely due to a small angular drift of the electron beam on a long time scale, resulting in a higher electron signal through the electron spectrometer pinhole. Therefore, only the short term variation of the charge can be estimated from this measurement, giving typical fluctuation of about $\mathrm{50\,pC/sr/shot}$ corresponding to 20\% RMS. 

Figure \ref{fig:stab}.b confirms the observations made previously regarding the stability of the spectrum, and indeed, shows that the mean energy of the electrons is quite stable at $\langle E\rangle\simeq\SI{2.1}{\mega\electronvolt}$ during the first 150\,min of monitoring, with short-term RMS variations of only 2-4\% (shaded area in Fig.~\ref{fig:stab}.b). The decrease of the mean beam energy  to $\langle E\rangle\simeq\SI{1.5}{\mega\electronvolt}$ toward the end of the run can also be clearly observed. Note that during the run, the laser system was extremely stable, see Fig.~\ref{fig:stab}.c, except for the energy measured using the turning box diagnostic (red curve in Fig.~\ref{fig:stab}.c), which is the last measurement point before the focusing parabola and is, therefore, the most representative of the evolution of the laser energy on target. The energy measured at this point decreased steadily during the experiment and reached a 11\% relative loss after 306\,min. It is likely that this progressive energy loss was due to the slow damage of a few chirped mirrors at the end of the compressor. The damage on the mirrors was found during the inspection which we performed the day after the experiment. Interestingly, the evolution of the electron energy can be correlated to the evolution of the laser energy at this last measurement point. To display these correlations, Fig.~\ref{fig:corr} shows the normalized mean energy of the electrons plotted against the laser relative energy. Two different correlation regimes are clearly distinguishable: (i) the first 5\% of laser energy loss leads to a $\mathrm{\sim}$ 5\% energy loss of the electrons suggesting a linear correlation. We then observe a threshold effect, (ii) as the next 5\% drop of laser energy correlates with a $\mathrm{\sim}$30\% electron mean energy loss. Assuming a linear dependence in both regimes, the two parts of the correlation plot are linearly fitted, yielding a slope $\mathrm{\alpha_1=1.0}$ in the first five percents of energy loss, and a slope $\mathrm{\alpha_2=5.6}$ in the following five percents.  This highlights the importance of laser energy stability: in our case, energy variations larger than  $5\%$ can cause significant modifications of the electron spectrum due to what seems like a threshold effect. Nevertheless, decent stability of the electron beam was achieved over the 300\,min of continuous operation, with the first 150\,min period displaying a remarkable stability correlated to the highest laser performance. Stability over multiple hours of operation with pC-MeV range electron beams at kilohertz repetition rate, corresponding to more than $\mathrm{18\times10^6}$\, consecutive shots, represents significant progress toward scientific applications of laser-plasma accelerators.     
\vskip 2mm

\begin{table}[ht!]
\begin{ruledtabular}
\renewcommand{\arraystretch}{1.25}
\begin{tabular}{c|c c c}
& Day 1 & Day 2 & Day 3\\
\colrule
I ($\mathrm{W.cm^{-2}}$) & $1.8\times10^{18}$ & $2.0\times10^{18}$ & $1.6\times10^{18}$ \\
\colrule
$\mathrm{n_{e,peak}}$ ($\mathrm{cm^{-3}}$) & $8.8\times10^{19}$ & $9.7\times10^{19}$ & $9.7\times10^{19}$\\
\colrule
Q (pC/shot) & 1.6$\mathrm{\pm}$0.2 & 2.6$\mathrm{\pm}$0.6 & 1.4$\mathrm{\pm}$0.2\\
\colrule
div. fwhm (mrad) & 42$\mathrm{\pm}$10 & 77$\mathrm{\pm}7$ & 57$\mathrm{\pm}$11\\
\colrule
$\mathrm{\langle E\rangle}$ (MeV) & 2.29$\mathrm{\pm}$0.13 & 2.11$\mathrm{\pm}$0.06 & 2.19$\mathrm{\pm}$0.04\\
\end{tabular}
\end{ruledtabular}
\caption{\label{tab:table1} Various experimental parameters and electron beam performance showing slight variations from day to day but overall fair reproducibility of the experiment. The values after the $\pm$ sign are RMS deviation.}
\end{table}

\subsection{\label{sub:reprod}Day to day reproducibility}

\begin{figure}[t!]
     \centering
         \includegraphics[width=8.5cm]{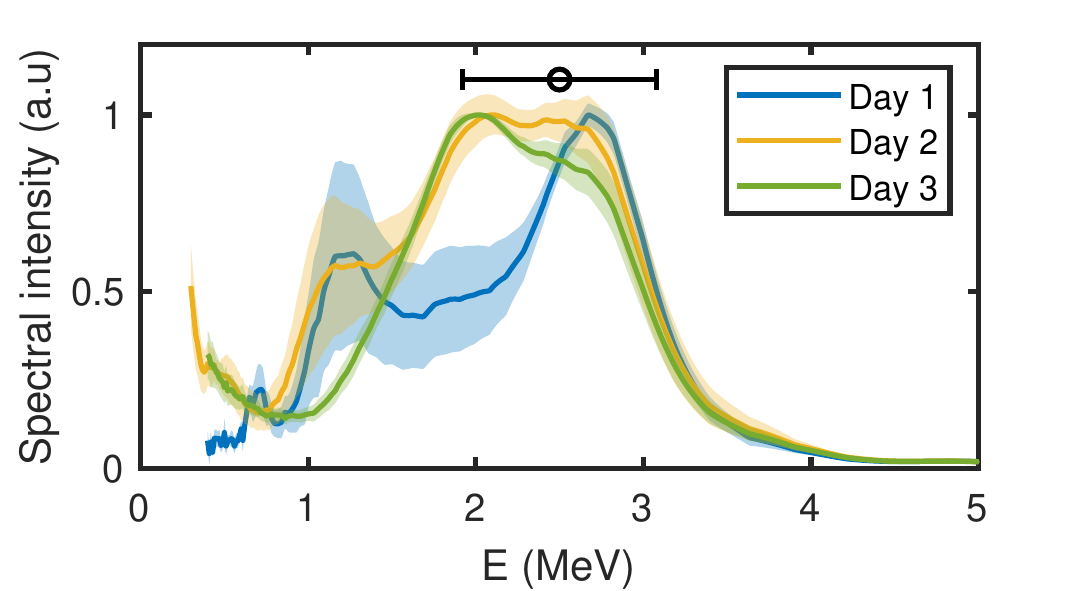}
        \caption{\label{fig:jours} Electron spectra obtained on three different days with the same one-sided shock nozzle. Spectra are the results of averaging over 2000\,shots for day 1 and day 2, and 5000\,shots for day 3. Day 1 and day 2 are 7 days apart, day 2 and day 3 are 6 days apart. }   
\end{figure}

In order to determine the reproducibility  of the electron beam and to assess the sensitivity of the accelerator to small day-to-day variations of the laser parameters, the measurements were repeated on three different days, each separated by about a week. The same actual OSS nozzle was used for the three experimental runs and showed no degradation over time. Table \ref{tab:table1} summarizes experimental conditions for each day, as well as the charge and mean electron energy corresponding to the electrons spectra displayed in Fig.~\ref{fig:jours}. These results show that the downward gradient injection method with one-sided shock nozzles increased significantly the reliability of the accelerator. Indeed, electrons beams with similar charge and 2-3\,MeV peaked spectrum were easily obtained even though the experimental parameters varied slightly from day to day. In particular, we see that experiments from day 2 and day 3, performed at the same plasma density, yield very similar electron spectra. This level of reproducibility was not observed in our previous experiments relying on ionization injection and self-injection. Such level of reproducibility is also decisive for a reliable use of the accelerator for applications. Moreover, these fused-silica nozzles showed a great resilience to damage, as the one used for this experiment provided reliable and reproducible results even after using it for about $\mathrm{30\text{\,-\,}50\times10^{6}}$\,shots. 



\section{\label{sec:conclusion}Conclusion}

In conclusion, by using an asymmetrically shocked fused-silica nozzle, we were able to produce the sharp density transition and the high plasma density necessary for downward density gradient injection in a LPA driven by few-cycle laser pulses. We managed to stabilize the injection and obtained electron beams with picocoulomb charge and MeV peaked spectrum for 5\,hours of continuous operation. In addition, the electron energy was correlated to the laser energy stability, showing that energy fluctuations should remain at the percent level in order to ensure stable accelerator performance. Finally, we demonstrated good day-to-day reproducibility over a period of two weeks thus making our accelerator ready for first application experiments.\\

\begin{acknowledgments}

This work was funded by the European Research Council (ERC Starting Grant FEMTOELEC) under Contract No. 306708. Financial support from the R\'egion \^Ile-de-France (under contract SESAME-2012-ATTOLITE) and the Extreme Light Infrastructure-Hungary Non-Profit Ltd (under contract NLO3.6LOA) is gratefully acknowledged. We also acknowledge  Laserlab-Europe, H2020 EC-GA 654148 and the Lithuanian Research Council under grant agreement No. S-MIP-17-79.
\end{acknowledgments}


\providecommand{\noopsort}[1]{}\providecommand{\singleletter}[1]{#1}%
\pagebreak
\widetext
\begin{center}
\textbf{\large Supplemental Materials}
\end{center}

\begin{figure*}[ht!]
     \centering
         \includegraphics[width=14cm]{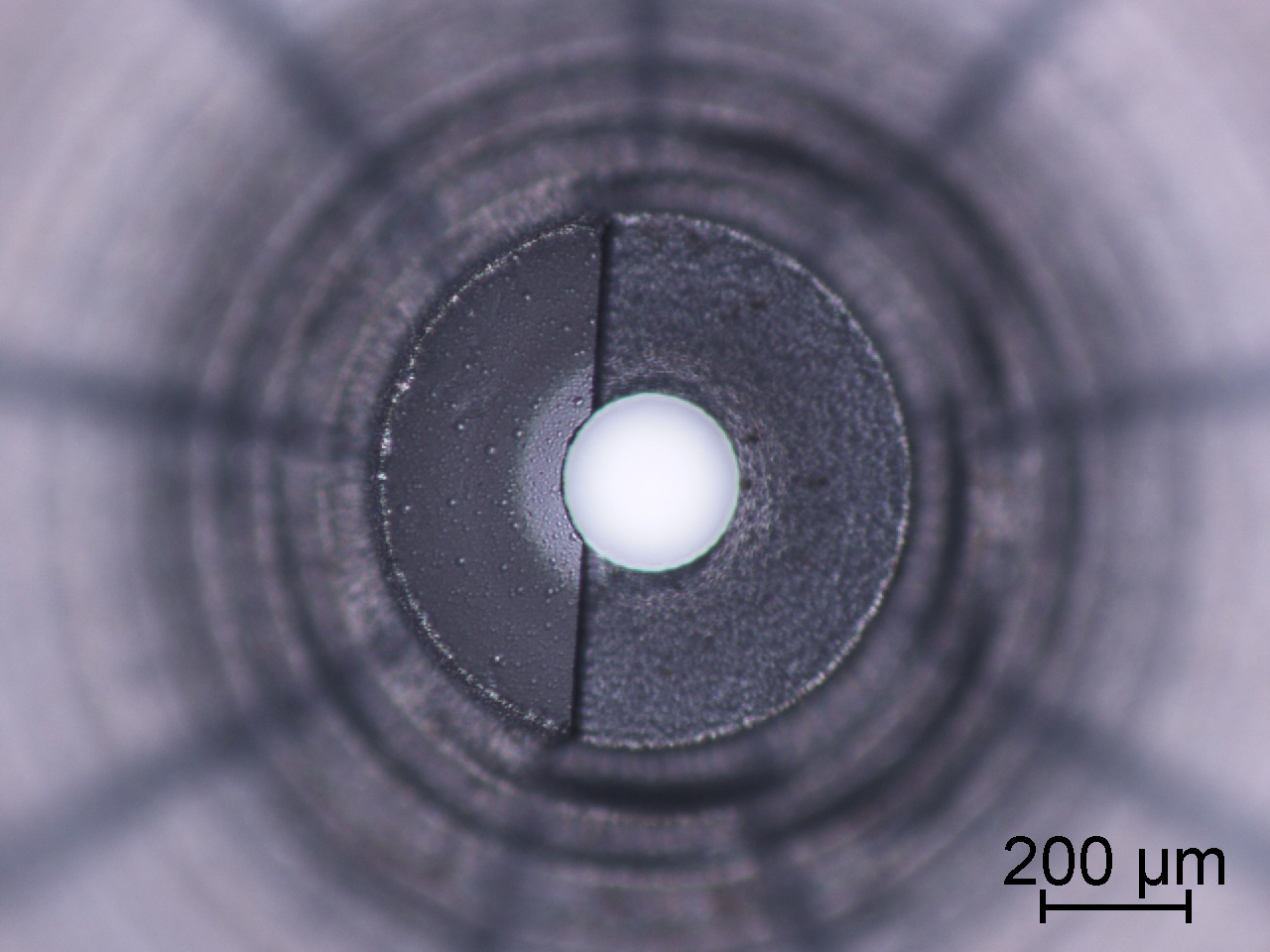}
        \caption{\label{fig:mic} Top-view of a one-sided shock nozzle taken with an optical microscope. }   
\end{figure*}

\end{document}